\documentclass[11pt]{article}
\usepackage{amsmath}
\usepackage{latexsym}
\usepackage{amssymb}

\numberwithin{equation}{section}
\title{From Concept to Reality to Vision
\footnote{Speech in acceptance of European Physical Society prize for
high energy physics, Aachen, August 2003.}}
\author{Frank
Wilczek\footnote{wilczek@mit.edu}}
\begin{document}
\maketitle

\begin{abstract}
I take a brief look at three frontiers of high-energy physics,
illustrating how important parts of our current thinking evolved from
earlier explorations at preceding
frontiers.
\end{abstract}
Elucidating the basic nature of the strong
interaction was a vast enterprise to which many gifted scientists
devoted their best efforts and made wonderful contributions. While
the subject is far from finished -- the dramatic developments I'll be
discussing this afternoon \cite{fw1} bring that home! -- I think it is clear
that the foundations are secure.  QCD, as the basic theory, is here
to stay. It is a marvelous theory, which cleanly embodies mathematical
ideas of great depth and beauty.  Above all QCD demonstrates, in a
unique way, the power of relativistic quantum field theory to produce
an amazing wealth of phenomena (asymptotic freedom, jets,
confinement, mass generation, resonance spectroscopy, chiral symmetry
breaking, anomaly dynamics, ...) in harmony with the observed facts
of Nature.

David Gross has just described for you the whirlwind of events and
discoveries that led us to propose this theory for the strong
interaction, reinforced with concrete reasons to believe in it (and
no other!), and packaged with proposals for critical, quantitative
experimental tests. I don't want to repeat the details, but only want
to endorse what David has already emphasized, that he and I were
fortunate indeed to be in a position to leverage a vast accumulation
of knowledge and technique built up by a big international community
of scientists over decades of dedicated work, much of it frustrating
and not properly recognized. As members of this community we should
all be proud of our joint achievement.

I'll freely admit that back in 1973 I didn't begin to anticipate the
progress in experiment and theory that would bring our subject to the
level where it is today. I had some hope that deep inelastic scattering
experiments and perhaps measurements of electron-positron
annihilation (the total cross section) would be made more precise,
maybe precise enough that with careful analysis one would see hints
of scaling deviations in the form we predicted, and thereby gradually
build up a case for the correctness of QCD.  Of course, reality has
far outrun these expectations.  One of the great joys of my life in
physics has been to participate in the process -- something like
parenthood -- whereby unshaped concepts mature in surprising ways
into concrete realities, which then engender new visions.  I'd like
briefly to share with you three examples, in each case mixing a
little nostalgia with pointers to the future.

\newpage

\section{From Running Coupling to Quantitative Unification to
\\ Supersymmetry}

Running of gauge theory couplings, and in
particular the peculiar anti-screening behavior we call asymptotic
freedom, was first established by straight unguided calculation 
\cite{dgross1973, polit1973}. It
was first applied to renormalization group equations for deep
Euclidean Green's functions and Wilson coefficients in operator
product expansions, enabled through a rather cumbersome formalism to
describe a very few physical processes \cite{dgrossb1973, dgross1974, 
hgeorgi1974}.  Before long antiscreening
was understood in terms accessible to intuition \cite{nielsen1981}. 
And by now of course
we've learned to exploit the concept much more boldly and
confidently, and with great success. Quark and gluon degrees of
freedom are identified directly in the energy-momentum flow of jets,
and their basic couplings are made manifest in the iconic three-jet
processes seen at LEP. Comparing the frequency of such events, at
different energies, exhibits the running in as clear and elementary a
form as one could ask for.

The calculation of running, of course, extends immediately to
electroweak interactions. (Indeed, my original interest in it largely
arose from this angle.) It was put to brilliant use in the famous
work of Georgi, Quinn, and Weinberg \cite{hgeorgib74}, who indicated 
through its use
dreams about unification of interactions (Pati and Salam 
\cite{pati1973}, Georgi and
Glashow \cite{fw10}) could be brought down to earth.  One could check concretely
whether the observed, unequal couplings might result from running a
single coupling from ultra-short to accessible distances. A few years
later Dimopoulos, Raby, and I  realized \cite{dim1981} (to my great surprise,
initially) that including the effects of low-energy supersymmetry,
which is quite a drastic expansion of the physics, makes only
comparatively small changes in the predictions that emerge from this
sort of calculation.   Precision experiments and improved
calculations appear to endorse these dreams and ideas, in their
supersymmetric version.

Unless this is a cruel tease on the part of Mother Nature, it means
we can look forward to a lot of fun exploring supersymmetry, and
maybe some aspects of unification, at the LHC. An especially poetic
possibility is to explore the possibility that other sorts of
parameters, besides gauge couplings, derive by running from a unified
value \cite{fw12}. It is widely speculated that the masses of 
different sorts of
gauginos, or of squarks and sleptons, might be related in this
way.

\section{From Dark Momentum to Gluonization to Higgs and Dark Matter}

Feynman interpreted the famous SLAC experiments on deep inelastic
scattering using an intuitive model of nucleons that postulated
point-like particles (partons) as nucleon constituents and treated
their dynamics in a crude impulse approximation, ignoring both
interactions and quantum interference \cite{feyn1970}. Identifying 
the partons as
quarks, and building the weak and electromagnetic currents by minimal
coupling to quarks, led to many successful predictions 
\cite{bjo1969}. There was,
however, one clear failure. The momentum carried by quarks inside a
fast-moving proton does not add up to the total momentum of the
proton, in fact it is less than half.

Today's ``dark matter" problem in astronomy is reminiscent of
this old ``dark momentum" problem. In the formal treatments of deep 
inelastic scattering, the analogy becomes eerily precise. In that
framework, the (failed) sum rule expresses the equality of the full
energy-momentum tensor with the energy-momentum tensor constructed
from quarks \cite{dgross1974, hgeorgi1974}.  Where electroweak 
currents see just quarks, gravitons
see more!  We realized early on \cite{dgross1974, hgeorgi1974} that 
the color gluons of QCD, which
are electroweak singlets but do carry energy-momentum, would enable
us to keep the good predictions while losing the bad one.  Evidently
the gluons had to be major, though ``dark" (or better: invisible),
constituents of the proton.

Our analysis of deep inelastic scattering, which followed pioneering
ideas of Wilson \cite{wil1971}, and built on the insightful hard work 
of Christ,
Hasslacher, and Mueller \cite{christ1972}, went beyond the parton 
model in other, more
profound ways. A fast-moving quark is revealed, to probes at higher
resolution (higher $Q^2$), to be composed of slower-moving (smaller
$x$) quarks, anti-quarks and gluons, which in turn will resolve into
more, softer stuff. This process, seen experimentally as evolution of
structure functions, is deeply characteristic of quantum field theory.

These evolution effects further enhance the role of glue in the
proton.  Several of us worked out that there should be a dramatic
pile-up of soft stuff, particularly soft glue, at small $x$ 
\cite{rujula1974}. To a
hard current (indirectly), or to a hard graviton (theoretically), the
proton mostly looks like a blob of soft glue. Twenty years later,
beautiful work at HERA confirmed these predictions in impressive
detail \cite{wolf}.

Very soft or ``wee" constituents of protons played a major role in
Feynman's ideas about diffractive scattering \cite{feyn1969}. His 
idea was that in
diffractive scattering, by exchange of wee partons, the relative
phases between different multiparton configurations in the proton
wave function get disrupted, without much transfer of
energy-momentum.   These ideas are intuitively appealing, and have
inspired some successful phenomenology, but as far as I know they
haven't yet been firmly rooted in QCD.

Much better understood -- I hope! -- is the importance of
gluonization for some frontier topics in high-energy physics, namely
Higgs particle production and WIMP searches.  The primary, classical
coupling of Higgs particles is to quarks, proportional to their mass.
But because the $u$ and $d$ quarks we mainly find inside nucleons are
so light, their direct coupling is heavily suppressed. Instead the
most important coupling arises indirectly, as a quantum effect,
through virtual top quark loops connecting to two gluons \cite{wilczek1977}.

I was originally interested in this Higgs-gluon vertex for its
potential to induce Higgs particle decays. Georgi, Glashow, Machacek, 
and Nanopoulos \cite{georgi1978} quickly
realized it could be exploited for production of Higgs particles, at
hadron colliders, through ``gluon fusion". This process, which of
course relies completely on the glue content of protons, is expected
to be the main production mechanism for Higgs particles at the LHC.
It is important to calculate the production rate accurately,
including good estimates of the gluon distribution functions, so that
we will be able to interpret the observed production rate, and check
whether the basic vertex is in fact what the standard model, in this
intricate way, predicts.

The on-shell Higgs particle couples to hard gluons. In its decay we
will see jets, and we can estimate the production using gluon
structure functions and perturbative QCD. When considering detection
of the sorts of dark-matter candidates provided by models of
low-energy supersymmetry we find ourselves involved in quite a
different kinematic domain. Since these WIMPs will be heavy and
slowly moving by particle physics standards, they will scatter at
very small momentum transfer. The coupling of SUSY WIMPs to depends
on poorly constrained details of the models, but in many realizations
it is dominated by virtual Higgs exchange. Here the Higgs-gluon
vertex comes in at essentially zero energy-momentum. Shifman,
Vainshtein and Zakharov \cite{shifman1978}, in beautiful work, 
related the relevant
gluon operator to the trace of the energy-momentum tensor, whose
matrix elements are of course known. This links back to the old dark
momentum problem, bringing us full circle.

It is philosophically profound, and quite characteristic of modern
physics, that even when viewing something so basic and tangible as a
proton, what you see depends very much on how you choose to look.
Low-energy electrons see point-like particles, the version described
in old high-school textbooks; hard currents see an evolving pattern
of quarks; gravitons see these plus lots of gluons as well; wee
gluons see some complicated stuff we don't properly understand (we do
know its name, Pomeron); real Higgs particles see gluons almost
exclusively; and WIMPS, through exchange of virtual Higgs particles,
see the Origin of Mass. (The trace of the energy momentum tensor, to
which they mainly couple, is on the one hand dominated by contributions from massless
color gluons and nearly massless quarks, and on the other hand equal to the nucleon mass.) 
Each probe
reveals different aspects of versatile reality.

\section{From Asymptotic
Simplicity to Quark-Gluon Plasma to Quark-Hadron Continuity}

I mentioned earlier how we've learned
to use the concept of asymptotic freedom more boldly and confidently
over the years.  To put it differently, we've learned fruitful ways
to lower our standards.   Instead of trying to prove directly from first principles
that weak coupling applies , we usually
content ourselves with consistency checks. That is, we tentatively
assume that perturbative calculation of some quantity of interest
starting with quark and gluon degrees of freedom is adequate, and
check whether the calculation contains infrared divergences 
\cite{sterman1977}. This
check is by no means trivial, since QCD is full of massless (color)
charged particles. So in cases where we find there are no infrared
divergences we declare a well-earned victory, and anticipate that our
calculations will approximate reality.  This strategic retreat has
licensed a host of successful applications to describe jet
processes, inclusive production, fragmentation, heavy quark physics,
and more.

We aren't always forced to compromise. In some important
applications, including low-energy spectroscopy, direct integration
of the equations using the techniques of lattice gauge theory is
practical. But as physicists hungry for answers, we properly regard
strict mathematical rigor as a desirable luxury, not an indispensable
necessity.

A particularly interesting and important application of the looser 
philosophy is to construct self-consistent descriptions of extreme 
states of matter, starting from quarks and gluons \cite{QCD2001}.

The high temperature, low baryon number regime is foundational for
very early universe cosmology.  It is also the object of an intense,
international experimental program in relativistic heavy ion physics.
The overarching theme is that a perturbative description of
high-temperature matter, starting with free quarks and gluons,
becomes increasingly accurate as the temperature increases. This can
be seen, for the equation of state, from numerical simulation of the
full theory \cite{skatz}. After heroic calculations, which introduce several
ingenious new techniques, controlled perturbative calculations
(including terms up to sixth order in the coupling, and some infinite
resummations) match the numerical work \cite{yschroder}. This is a milestone
achievement in itself, and also promising for future developments,
since the perturbative techniques are more flexible. They might be
applied, for example, to calculate viscosity and energy loss, which
can be probed experimentally. In this way, we can hope to do justice
to the vision of quark-gluon plasma.

The regime of high baryon number density, and low temperature, is
intrinsically fascinating, and might be important for describing the
inner dynamics of supernovae and the deep interior of neutron stars.
The first fundamental result about QCD at high baryon number density
is that many of its key properties, including for example the
symmetry of the ground state and the energy and charge of the
elementary excitations, can {\it not\/} be calculated to a good
approximation starting from fermi balls of non-interacting quarks.
The perturbation theory (for just about anything) contains infrared
divergences \cite{QCD2001}.

Fortunately, the main source of these divergences is well understood.
It signals an instability toward the development of a condensate of
quark pairs, similar to the Cooper pairs that occur in metallic
superconductors. Whereas the phenomenon of superconductivity in
metals is very delicate, because one must overcome the dominant
Coulomb repulsion of like charges, color superconductivity is very
robust, because there is a fundamentally attractive force between
quarks (in the color and flavor antitriplet, spin singlet channel).
One can construct an approximate ground state that accommodates the
pairs, using the methods of BCS theory. Perturbation theory around
this new ground state no longer has infrared divergences. Thus we
find that strongly interacting matter at asymptotically high density
can be studied using weak coupling, but non-perturbative methods.

Color superconductivity has become an extremely active area of
research over the past few years, and many surprises have emerged.
Perhaps the most striking and beautiful result is the occurrence of
color-flavor locking, a new form of symmetry breaking, in real-world
(3 flavor) QCD at asymptotic densities \cite{malford1999}. The symmetry
$SU(3)_C \times SU(3)_L \times SU(3)_R$ of local color times chiral flavor is broken
down to the diagonal subgroup, a residual global $SU(3)$.

Color-flavor locking is a rigorous, calculable consequence of QCD at
high density. It implies confinement and chiral symmetry breaking.
The low-energy excitations are those created by the quark fields,
those created by the gluon fields, and the collective modes
associated with chiral symmetry breaking. Because CFL ordering mixes
up color and flavor, the quarks form a spin-1/2 octet (plus heavier
singlet), the gluons form a vector octet, and the collective modes
form a pseudoscalar octet under the residual $SU(3)$.
Altogether there is an uncanny resemblance between the properties of
dense hadronic matter one calculates for the CFL phase, and the
properties one might anticipate for ``nuclear matter" in a world with
three massless quarks.
A nice perspective on this arises if we consider coupling in the U(1)
of electromagnetism. Both the original color gauge symmetry and the
original electromagnetic gauge symmetry are broken, but a combination
survives. This is similar to what happens in the standard electroweak
model, where both weak isospin and hypercharge are broken, but a
certain combination survives (to provide electromagnetism). Just as
in that case, also in CFL+QED the charge spectrum is modified. One
finds that the quarks, gluons, and pseudoscalars acquire integral
charges (in units of the electron charge); in fact, the charges
precisely match those of the corresponding hadrons.

It is difficult to resist the conjecture that these two states are
continuously related to one another, with no phase transition, as the
density varies \cite{schaef1999}.  During this variation, degrees of 
freedom that are
``obviously" three-quark baryons evolve continuously into degrees of
freedom that are ``obviously" single quarks. This nifty trick is
possible because diquarks can be exchanged freely with the condensate.

If the core of a neutron star is described by the color-flavor locked
(CFL) phase, which seems plausible, it will be a transparent
insulator that partially reflects light -- like a diamond! This
particular consequence of the CFL phase is unlikely to be observed
any time soon, but we are working toward defining indirect signatures
in observable neutron star and supernova properties.

Unfortunately, existing numerical methods for calculating the
behavior of QCD converge very slowly at high density and low temperature. They
are totally impractical, even for the biggest and best modern
computers.    Developing usable algorithms for this kind of problem
is a most important open challenge.

\bigskip
\bigskip

There are other stories linking the past with the future through
asymptotic freedom and QCD, including a particularly interesting and
potentially important one involving axions.  But I'll stop here.
Thanks again.

  \end{document}